\documentclass[twocolumn,showpacs,preprintnumbers,amsmath,amssymb]{revtex4}
\usepackage{graphicx}
\begin{document}
\def\T{\Theta}
\def\D{\Delta}
\def\d{\delta}
\def\r{\rho}
\def\p{\pi}
\def\a{\alpha}
\def\g{\gamma}
\def\ra{\rightarrow}
\def\s{\sigma}
\def\b{\beta}
\def\e{\epsilon}
\def\G{\Gamma}
\def\om{\omega}
\def\pe{$1/r^\a$ }
\def\l{\lambda}
\def\f{\phi}
\def\w{\psi}
\def\m{\mu}
\def\dg{d^3{\bf r}\,d^3{\bf p}}
\def\df{f({\bf r, p})}
\def\dn{n({\bf r, p})}
\def\be{\begin{equation}}
\def\ee{\end{equation}}

\title{Phase diagram of self-attracting systems}

\author{P.H.~Chavanis$^{1}$ and I.~Ispolatov$^{2}$}
\affiliation{$^{1}$Laboratoire de Physique Quantique, Universit\'e Paul Sabatier, 118 
route de Narbonne 31062 Toulouse, France\\
$^{2}$Departamento de Fisica, Universidad de Santiago de Chile,
Casilla
302, Correo 2, Santiago, Chile}
  
\begin{abstract}

Phase diagram of microcanonical ensembles of self-attracting particles
is studied for two types of short-range potential regularizations:
self-gravitating fermions and classical particles interacting via
attractive soft $-(r^2+r_0^2)^{-1/2}$ Coulomb potential.  When the
range of regularization is sufficiently short, the self-attracting
systems exhibit gravitational or collapse-like transition.  As the
fermionic degeneracy or the softness radius increases, the
gravitational phase transition crosses over to a normal first-order
phase transition, becomes second-order at a critical point, and
finally disappears.  Applicability of a commonly used saddle-point or
mean-field approximation and importance of metastable states is
discussed.

\end{abstract}

\maketitle

\section{Introduction}
\label{sec_intro}

Ensembles of particles interacting via a long-range nonintegrable
attractive potential, $U(r)=A r^{-\a}$, $A<0$, and $0<\a<3$, are known
to exhibit gravitational phase transition between a relatively uniform
high energy state and a low-energy state with a core-halo structure
\cite{ll,lbw,ki,pr,ki2,ch1,bm2,usg,dv,ch,chs,aa}.  It has
also been established \cite{pr,ki2,ch1,bm2,usg,ch,dv} that if the
interaction potential has some form of short range cutoff, the density
and all other physical quantities that characterize the core-halo
state are finite, while for a bare $r^{-\a}$ attractive potential the
collapse results in a singular state with an infinite density, entropy
and free energy.  The gravitational phase transitions are known to
exist in microcanonical, canonical, and grand canonical ensembles; yet
the details of the phase transition and the structure of the core-halo
state are ensemble-dependent.  In this paper we shall essentially
consider microcanonical ensembles; for long-range interacting systems
they are known to have richer phenomenology and allow for states (such
as ones with a negative specific heat) that are inaccessible in both
canonical and grand canonical ensembles. In addition, the
microcanonical ensemble is the most fundamental since the notion of
thermostat is ambiguous for long-range interacting systems.

\begin{figure}
\includegraphics[width=.5\textwidth]{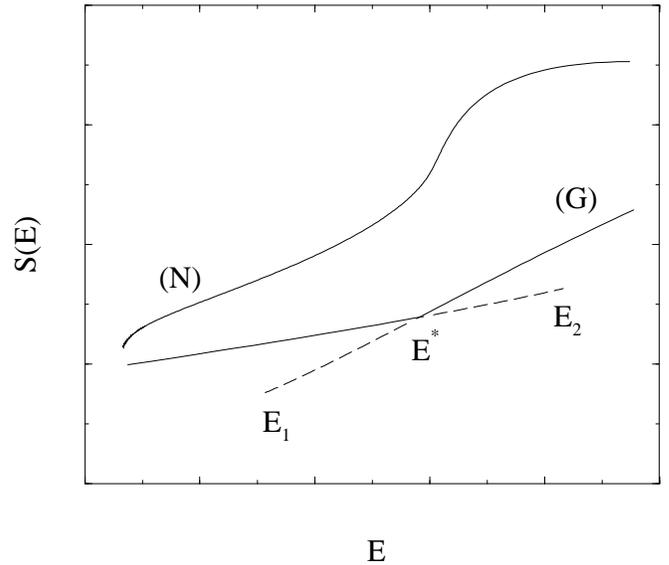}
\caption{\label{ESintro}
Sketch of an entropy vs. energy plot for a microcanonical
self-attracting system with a short-range cutoff which exhibits a
gravitational $(G)$ phase transition. Stable states and metastable
states are shown by solid and dashed lines, respectively.  Points
$E_1$, $E^*$, and $E_2$ denote the collapse energy for the uniform
state, the energy of the phase transition, and the energy for which
the metastable core-halo state becomes unstable, correspondingly.  A
typical entropy vs. energy plot for a microcanonical long-range
interacting system with a normal $(N)$ first-order phase transition is
also shown.  }
\end{figure}

A typical entropy vs. energy plot, describing a
gravitational phase transition in a microcanonical ensemble, is shown in
Fig. \ref{ESintro}, lower plot \cite{bm2,ki2,usg,ch1,ch}.
High-energy and low-energy branches correspond to the uniform and to the
core-halo states, respectively; their intersection $E^*$ marks the point
of the phase transition. Metastable uniform and core-halo states,
shown by dashed lines, exist in the energy intervals $[E_1,E^*]$ and
$[E^*, E_2]$, correspondingly.  When the energy is larger than $E_2$,
the system can be only in the stable uniform state.  The system still
stays in this uniform state after energy is decreased past $E^*$ and
the uniform state becomes metastable. When the energy reaches $E_1$,
the uniform state becomes unstable and the system undergoes a collapse
to a stable core-halo state. During such a collapse the entropy is
discontinuously increased, and the macroscopic rearrangement of the
density profile occurs.  Similarly, the core-halo state is stable
below $E^*$, metastable between $E^*$ and $E_2$, and undergoes a
discontinuous transition reverse to the collapse, at $E_2$. This
transition is sometimes called an explosion \cite{ki2}, since it
transforms the dense core into a relatively uniform mass distribution.

In some sense, the way the gravitational phase transition occurs in a
microcanonical ensemble resembles a hysteresis phenomenon that takes
place during a first-order phase transition in a canonical ensemble.
For this reason, it is sometimes called a {\it gravitational} first
order phase transition. Yet in a microcanonical ensemble a {\it normal}
first-order phase transition occurs without the hysteresis and
metastable states. A distinct feature of a normal microcanonical
first-order phase transition in a small or long-range interacting system 
is a convex dip in otherwise concave continuous entropy plot (see, for
example, \cite{gr,hu,us}) sketched in Fig. \ref{ESintro}, upper plot $(N)$.
Differences between gravitational and normal first-order phase
transitions are evident immediately: for a gravitational phase
transition the microcanonical inverse temperature $\b\equiv dS/dE$ is
discontinuous at $E^*$ while for the normal first-order phase
transition $dS/dE$ is always continuous. In addition, in a
gravitational phase transition, the uniform and the core-halo phases
cannot coexist, while for the normal first-order phase transition the
phases do coexist in a range of energies corresponding to the convex
dip.

It has been noticed \cite{bm2,ch1,usg,ch,ki2} that the gravitational phase
transitions in self-attracting systems exist only when the range of
the cutoff is sufficiently small.  When the effective cutoff radius is
increased, the range of existence of metastable states $[E_1,E_2]$
shrinks and finally disappears along with the $dS/dE$ discontinuity at
$E^*$. The resulting entropy vs. energy plot becomes continuous and
qualitatively resembles a corresponding plot for a system with a
normal first-order phase transition \cite{ch,usg}.  This was observed
in self-gravitating systems with various short-range regularizations:
central excluded volume
\cite{bm2}, hard sphere repulsion for individual particles \cite{ki2},
soft-core interaction potential \cite{usg}, and exclusion due to
Fermi-Dirac statistics \cite{ch1,ch}. 

In this paper we study how the behavior of self-attracting systems
depends on the effective short-range cutoff radius.  We find that once
the cutoff is increased above a certain value, the gravitational phase
transition crosses over to a normal first-order phase transition,
characterized by a convex dip in the entropy vs. energy plot and
associated with an energy interval with a negative specific heat.  As
the effective cutoff radius is increased even further, the system
reaches a critical point where the first-order phase transition is
replaced by a second-order one, and for even larger cutoff radii there
is no phase transition at all.  To reveal that such a phase diagram is
a generic feature of all self-attracting systems, we consider two
different examples: an ensemble of self-gravitating particles with
Fermi-Dirac exclusion statistics, and a particle system with classical
statistics interacting via a soft Coulomb potential
$-(r^2+r_0^2)^{-1/2}$.  Since both of these examples have been studied
in the past, in the following two sections we present only short
outlines of the derivation of the main formulas and refer readers to
\cite{dv,ch, ch1,usg} for a more detailed description.  After
analyzing these two examples, we discuss the validity of the
saddle-point or mean-field approximation, which is commonly used to
study self-gravitating systems; a conclusion which summarizes the
presented results completes the paper.

\section{self-gravitating fermions}
\label{sec_fermions}

We consider a microcanonical ensemble of $N\gg1$ identical
unit mass fermions confined to a spherical container of radius
$R$.  A simple combinatorial analysis gives for the number of
microscopic states $W$ of a Fermi-Dirac system with a phase space
distribution function $\df$:
\begin{eqnarray}
\label{ns}
\nonumber
W[f]\approx\\
\exp \left\{-{g\over (2\pi\hbar)^3}
\int[ n \ln n + (1-n) \ln (1-n) ] \dg \right\}.
\end{eqnarray}
Here $\dn\equiv(2\pi\hbar)^3\df/g$; $g$ is a number of internal
degrees of freedom, $g=2s+1$ for a particle with spin $s$. The
logarithm of Eq. (\ref{ns}) is the Fermi-Dirac entropy. The most
probable distribution expected at equilibrium is obtained by
maximizing Eq.~(\ref{ns}) with respect to $n$ with the constrained
total energy $E$ and number of particles $N$. The validity of this
mean-field approach is discussed in Sec. \ref{sec_mf}. Expressing the
particle density through the Laplacian of the gravitational field by
using the Poisson equation, it is found that the mean-field
gravitational potential $\f({\bf r})$ satisfies an equation of the
form
\be
\label{p}
\Delta\f({\bf r})={4\p Gg\over (2\p\hbar)^3} \int{d^{3}{\bf p}\over 1+
e^{[p^2/2+\f({\bf r})-\nu]/T}},
\end{equation}
where $G$ is the gravitational constant. The temperature $T(E,M)$ and
the chemical potential $\nu(E,M)$ appear as Lagrange multipliers
associated with the conservation of energy and mass. 
We assume a spherical symmetry of the
mean-field potential, introduce a dimensionless radius $\xi\equiv
r(Gg\sqrt{8 T}/\p\hbar^3)^{1/2}$, a non-negative dimensionless
potential $\w(\xi)\equiv[\f(r)-\f(0)]/T$, and a positive constant
$k\equiv\exp\{[\f(0)-\nu]/T\}$. Then Eq.~(\ref{p}) is reduced to
the form \cite{ch1}:
\be
\label{pd}
{d^2\w\over d\xi^2}+{2\over \xi}{d\w\over d\xi}=
I_{1/2}[ke^{\w(\xi)}].
\end{equation}
Here $I_{1/2}$ denotes the Fermi integral
\be
\label{fi}
I_{\sigma}(t)=\int_0^{+\infty}{x^{\sigma}\over 1+t e^x } dx.
\end{equation}
Boundary conditions $\w(0)=0$ and $\w'(0)=0$ follow from the
definition of the potential $\w(\xi)$ and spherical symmetry,
respectively.  Using the Gauss theorem for the gravitational potential
at the boundary of the spherical container $R$, the number of
particles constraint can be expressed as
\be
\label{bc}
{d\w\over  dr}(R)={GM\over T R^2}.
\end{equation}
Introducing a dimensionless maximum radius
$\a\equiv R(Gg\sqrt{8T}/\p\hbar^3)^{1/2}$, we reduce
Eq.~(\ref{bc}) to the form
\be
\label{bd}
\a^5\w'(\a)=\mu^2,
\end{equation}
where $\mu\equiv\sqrt{8g^2G^3MR^3/\p^2\hbar^6}$ is a degeneracy
parameter \cite{ch1}, which is proportional to the ratio of the
gravitational Fermi energy (defined in Eq. (\ref{bfer}) below) to the
average gravitational energy $GM^{2}/R$ to the power 3/2. In the
large-$\mu$ limit the system becomes completely non-degenerate or
classical, and Eq. (\ref{p}) then reduces to the Boltzmann-Poisson
equation. Yet for any finite $\m$ and sufficiently low energy 
the system is degenerate and mimics the  structure of a  white dwarf 
star \cite{chandr}. 
In the completely degenerate limit $T\rightarrow 0$, the equation for
the gravitational potential becomes the Lane-Emden equation
for a polytrope of index $n=3/2$ \cite{chandr}. The density goes to
zero at a finite radius $R_{*}$ related to the total mass $M$ by the
relation $MR_{*}^{3}=91.869\hbar^{6}G^{-3}(2/g)^{2}$.  The energy of
this fermion ball is given by the Ritter's formula
$E=-(3/7)GM^{2}/R_{*}$. Hence the minimal possible energy of 
a self-gravitating degenerate gas, i.e. the gravitational Fermi energy is
\be
\label{bfer}
E_{min}=-5.98\ 10^{-2} g^{2/3}{G^{2}M^{7/3}\over\hbar^{2}}.
\end{equation}
Note that contrary to the completely degenerate ($E=E_{min}$) case,
the density of the self-gravitating Fermi gas for $E>E_{min}$
decreases like $r^{-2}$ at large distances, similarly to the classical
self-gravitating isothermal gas. Hence the container is necessary to confine
the system \cite{bt}.

\begin{figure}
\includegraphics[width=.4\textwidth]{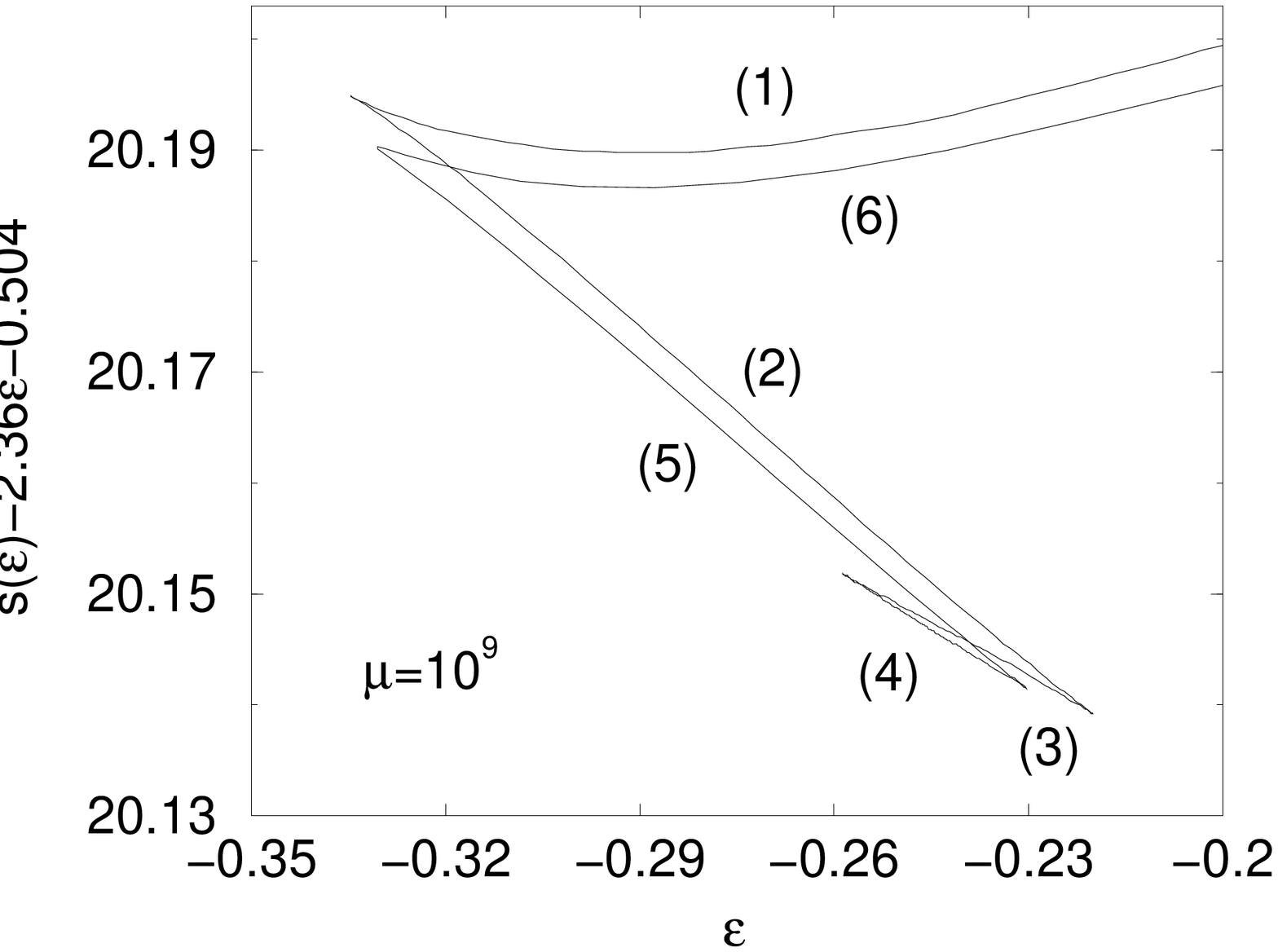}
\includegraphics[width=.4\textwidth]{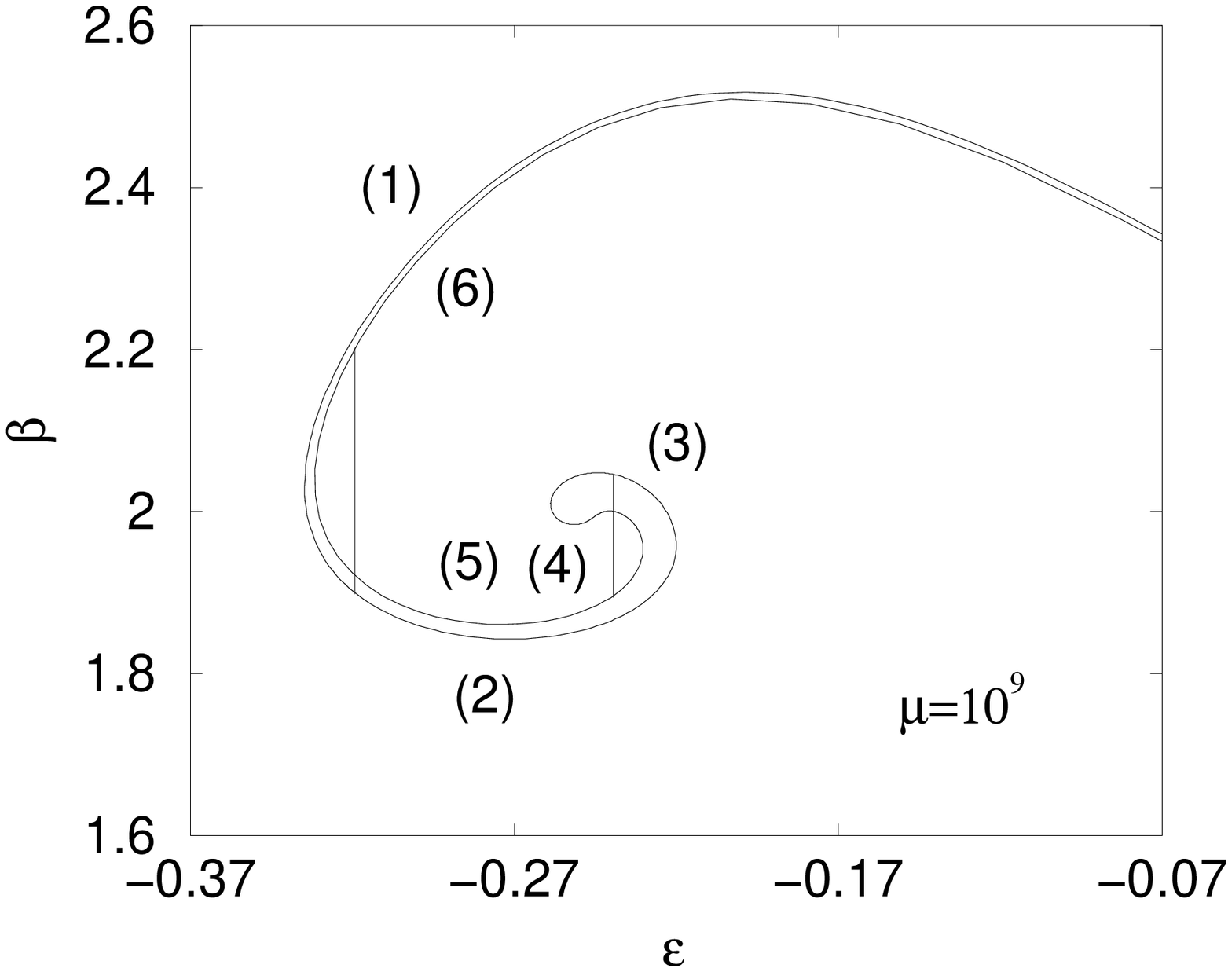}
\caption{\label{fignew}
Plots of entropy per particle $s(\e)$ (upper plot) and dimensionless
inverse temperature $\b(\e)=ds/d\e$ (lower plot) for $\m=10^9$. A
change of stability of the corresponding state occurs each time the
$(\e,\b)$ spiral has a vertical tangent and the $(\e,s)$ plot has a
cusp.  A mode of stability is lost at the vertical
tangent if the curve rotates anticlockwise and gained if the curve rotates 
clockwise \cite{ka1}. Therefore, the branches $(1)$ and $(7)$
are stable, $(2)$ and $(6)$ are unstable against one mode, $(3)$ and
$(5)$ against two modes and $(4)$ against three modes (branch (7) is not represented 
in the figure but is the continuation of branch (6) after the turning point of energy 
at $\epsilon\simeq 147$). There exists
values of energy at which the branches with the same degree of
instability have the same entropy. It occurs at the crossing points of
the $s$ vs $\e$ plot and corresponding vertical lines in the $\beta$
vs $\e$ plot. As $\mu\rightarrow +\infty$, there are more and more
crossing points at energies $\epsilon_{n}$ converging to the value
$\epsilon=1/4$ of the singular isothermal sphere. These points
$(\e_n,s_n)$ can be associated with points of phase transitions;
however, these ``phase transitions'' occur between unstable states and
are therefore unphysical.
}
\end{figure}
\begin{figure}
\includegraphics[width=.45\textwidth]{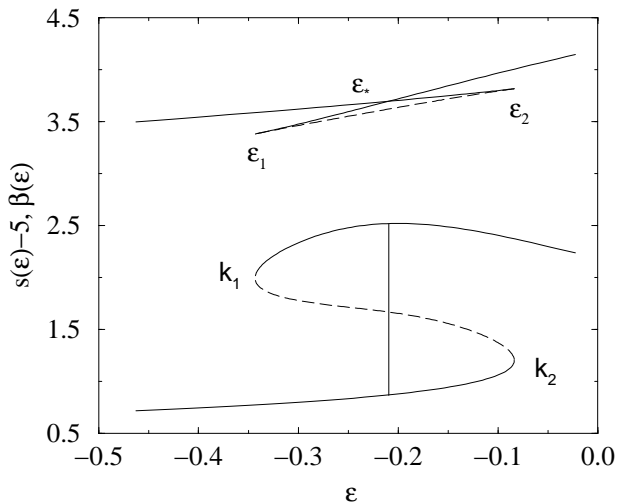}
\caption{\label{EbetaSMU10e4}
Plots of entropy per particle $s(\e)-5$ (top) and dimensionless inverse
temperature $\b(\e)=ds/d\e$ (bottom) for $\m=10^4$. Unstable state is
shown by dashed lines.  }
\end{figure}
\begin{figure}
\includegraphics[width=.45\textwidth]{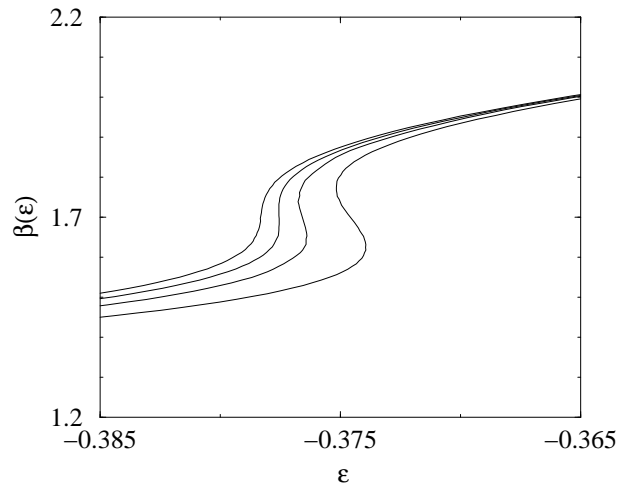}
\caption{\label{multimu1}
Plots of inverse
temperature $\b(\e)=ds/d\e$
for (left to right) $\m=2.65\times 10^3,\;2.67\times 10^3,\;2.7\times 10^3$,
and $2.75\times 10^3$. The gravitational phase transition disappears for 
$\mu<\mu_{gr}=2.67\times 10^3$ and is replaced by a normal first order phase 
transition.
}
\end{figure}
\begin{figure}
\includegraphics[width=.45\textwidth]{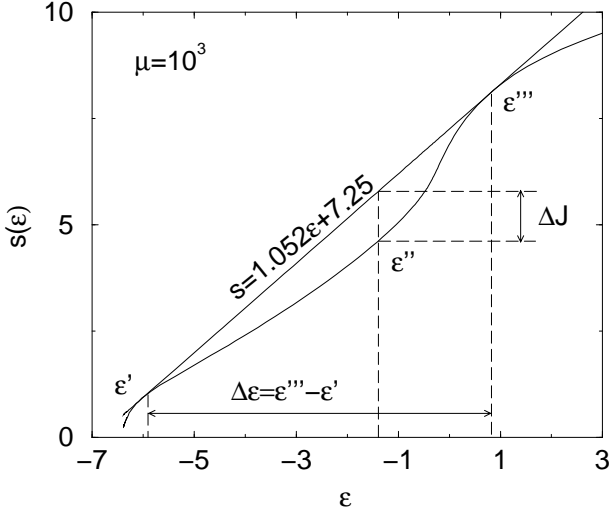}
\caption{\label{maxwell1} Plot of entropy $s$ vs. energy $\e$ for
$\m=10^3$. The entropy presents a convex intruder between
$\epsilon'$ and $\epsilon'''$. For an extensive system, this
convex intruder is forbidden because the system with energy
$\bar \e$, $\e'<\bar \e<\e'''$, would gain entropy by
splitting in two phases with energy $\epsilon' $ and
$\epsilon''' $. 
Indeed,
$s(\overline{\epsilon}=\alpha \epsilon'+(1-\alpha) \epsilon''')\le
\alpha s(\epsilon')+(1-\alpha) s(\epsilon''')$, where
$0\le\alpha \le 1$ parametrizes the energy $\bar \e$ of the systems
in the phase coexistence range $[\e',\e''']$.
However, for a non-extensive system, 
such as a gravitational
system, this argument does not hold and a convex intruder for the
entropy is allowed in the microcanonical ensemble \cite{gr}.} 

\end{figure}
\begin{figure}
\includegraphics[width=.45\textwidth]{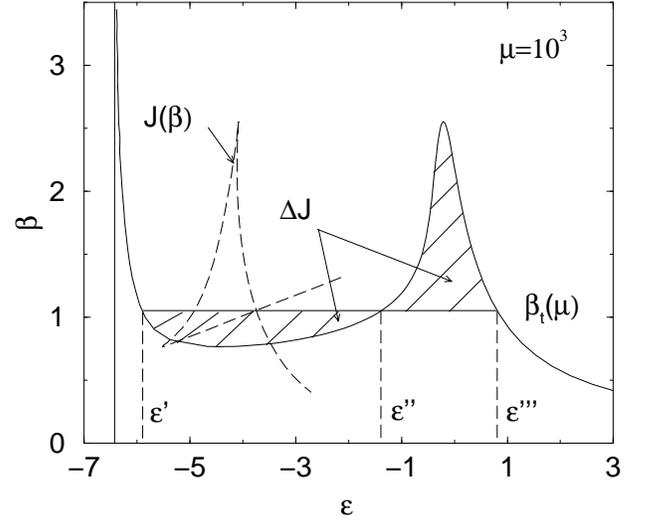}
\caption{\label{maxwell2} Plot of inverse
temperature $\b(\e)=ds/d\e$ vs. energy $\e$ for $\m=10^3$.  The
existence of negative specific heats $C=-\beta^{2}d\epsilon/d\beta<0$
and the convex intruder for the entropy are the signal of a normal
first order phase transition and of the {\it inequivalence} of
statistical ensembles. Indeed, for nonextensive systems, the region of
negative specific heat is allowed in the microcanonical ensemble while
it is forbidden in the canonical ensemble and replaced by a sharp
phase transition (horizontal plateau). The temperature of the
transition $\beta_{t}^{-1}$ is determined by the crossing point in the
free energy $J=s-\beta\epsilon$ vs inverse temperature $\beta$ plot
(dashed line). Alternatively, it can be obtained by performing a
Maxwell construction in the $\beta$ vs $\epsilon$ diagram, noting that
$\int_{\epsilon_{1}}^{\epsilon_{3}}(\beta-\beta_{t})d\epsilon=J'''-J'=0$
(the areas of the shaded regions are $-\Delta J=J''-J'$ and $\Delta
J=J'''-J''$). It is also given by the slope of the straight line
$s_{hull}(\epsilon)$ in Fig. \ref{maxwell1} ($s_{hull}=a\epsilon+b$
with $a=\beta_{t}$ and $b=J'=J'''$) \cite{gr}. During the
canonical phase transition, a latent heat
$\Delta\epsilon=\epsilon'''-\epsilon'$ is released.}
\end{figure}
\begin{figure}
\includegraphics[width=.45\textwidth]{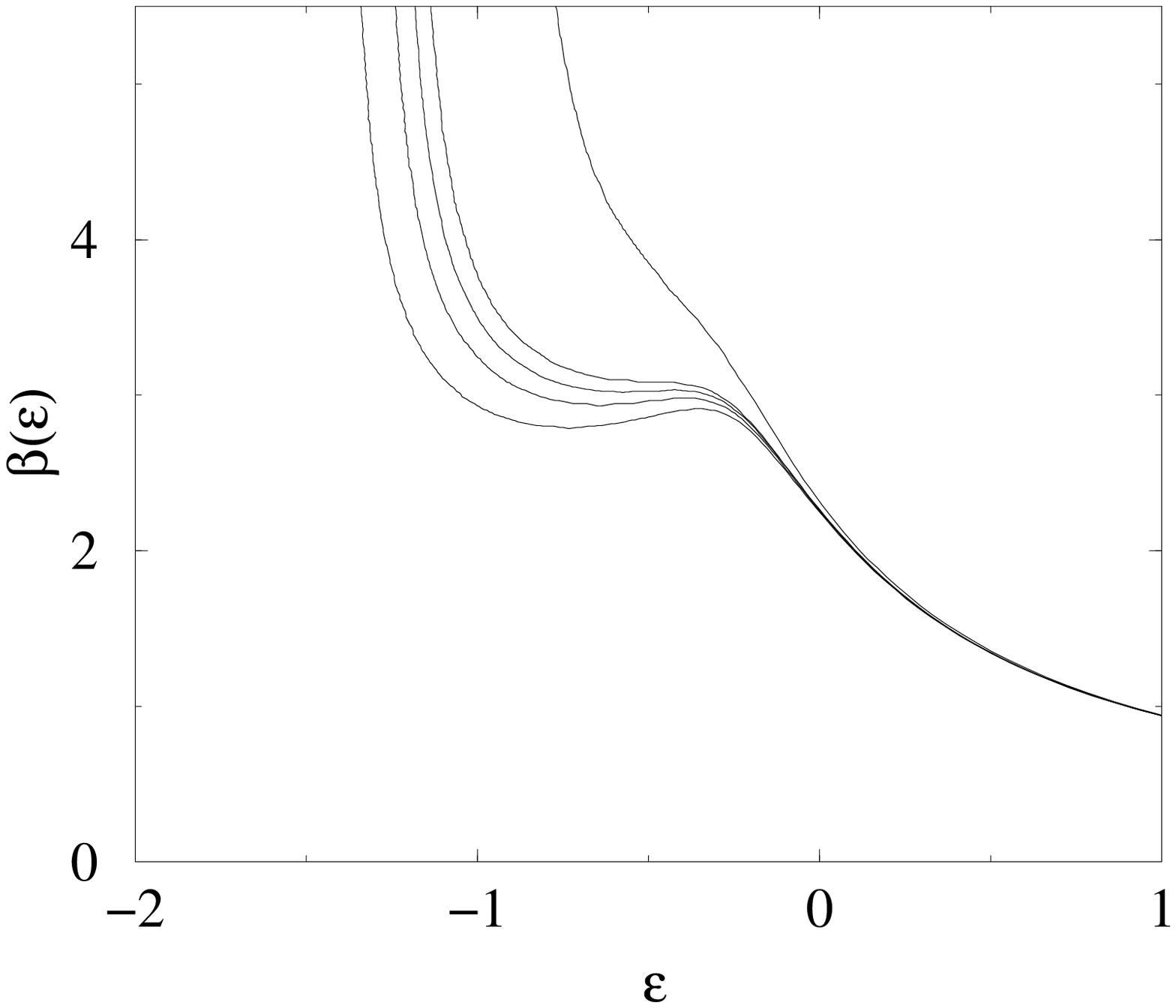}
\caption{\label{multimu2}
Plots of inverse
temperature $\b(\e)=ds/d\e$ vs. energy $\e$
for (left to right) $\m=100,\;90,\;85,\;80,$
and 50. The normal first order phase transition disappears for 
$\mu\le\mu_{cr} \approx 83$. At the critical value $\mu=\mu_{cr}$ the 
system passes by a second order phase transition.}
\end{figure}

As in  \cite{ch1}, we introduce a dimensionless energy $\e$, a dimensionless
inverse temperature $\b$, and the entropy per particle $s$, and
express them through $\a$, $\m$, $\w$ and $k$ as 
\begin{eqnarray}
\label{def}
\nonumber
\e\equiv{ER\over GM^2}=-{\a^7\over\m^4}\int_0^{\a}I_{3/2}[ke^{\w(\xi)}]\xi^2
d\xi \\
\nonumber
+{2\a^{10}\over3\mu^4}I_{3/2}[ke^{\w(\xi)}],\\
\nonumber
\b\equiv{GM\over RT}={\mu^2\over\a^4}, \\
\nonumber
s\equiv{\ln W\over M}={7\over 3}\e\b+\w(\a)+\b+\ln k\\
-{2\a^6\over 9\m^2}I_{3/2}[ke^{\w(\a)}].
\end{eqnarray}
Hence, all the thermodynamics parameters $\e$, $s$, and $\b$ are
single-valued functions of the degeneracy parameter $\m$ and
the uniformization variable $k$, which continuously parametrizes the
functions $\w(\xi)$.  
We select a degeneracy $\m$, and for each $k$ we
numerically integrate Eq.~(\ref{pd}) for $0\leq \xi \leq \a$, where
$\a$ is determined by Eq.~(\ref{bd}).  Once $\w(\xi)$ is computed,
$\e(k)$, $\b(k)$, and $s(k)$ are calculated using Eq.~(\ref{def}).  The
curves giving the entropy $s(k)$ vs. energy $\e(k)$ and the inverse
temperature $\b(k)=ds/d\e$ vs. energy $\e(k)$ are thus defined in a
parametric form for various values of the degeneracy parameter $\m$
(see Figs. \ref{fignew}-\ref{multimu2}).

In the classical $\mu\rightarrow +\infty$ limit, 
the $(\e,\b)$ plot forms a spiral which winds up indefinitely 
around a
limit point corresponding to the singular isothermal sphere
with the energy $\e=-1/4$ and inverse temperature 
$\b=2$ \cite{bt}. 
For very high but finite values of $\m$, such as $\m=10^9$ in 
Fig.~\ref{fignew},
the system is almost non-degenerate and its $\b$ vs. $\e$ plot looks
very similar to this infinite winding spiral.  
Yet as $\m$ is finite, the spiral is
finite as well and does unwind after a turning point. Each point with
a vertical tangent (where $d\b/d\e=\infty$) on the $\b$ vs. $\e$
spiral corresponds to a cusp in the $s$ vs. $\e$ plot.  However, all
the exotic features of the high-$\m$ plots, such as multiply-winding
spirals  and corresponding multiple kinks 
are related to unstable and
therefore physically unrealizable states \cite{ka1}.

As the degeneracy parameter $\mu$ is decreased, the number of
unstable states is decreased as well.
As illustrated in 
Fig.~\ref{EbetaSMU10e4}, for $\m=10^4$ there is only one 
unstable state, shown on the plot by a dashed line. 
Yet for $\m=10^4$ the system is still highly 
non-degenerate and
exhibits all signs of the gravitational phase transition.  The
low-energy branch of the plot corresponds to the core-halo state which
exists for $0<k<k_2$ and $\e_{min}<\e<\e_2$, where $\e_{min}=-6.42\
10^{-2}\mu^{2/3}$, is the dimensionless gravitational Fermi
energy.  Numerical evidence suggests that at the $\e_2$-cusp, both
energy $\e(k)$ and entropy $s(k)$ behave as ${\cal O}(k-k_2)^2$, which
explains the divergence of $d\b/d\e$, evident in the plot.  For
$k_2<k<k_1$ the entropy $s(k)$ is not even a local maximum
\cite{ch1,ch} so the corresponding state, marked by the dashed line
in Fig. \ref{EbetaSMU10e4}, is unstable.  When the curve approaches
the second cusp at $\e_1=\e(k_1)$, both energy $\e$ and entropy $s$ go
through a ${\cal O}[(k-k_1)^2]$ asymptotics again. For $k>k_1$ the
equilibrium states, now belonging to the uniform phase, are at least
locally stable.

When $\m$ gets smaller and the degeneracy becomes more important, the
interval between $\e_1$ and $\e_2$ decreases and finally disappears.
In Fig.~\ref{multimu1} we plot $\b$ vs. $\e$ and seek $\mu$ such that the
$\b(\e)$ curve loses its characteristic for large $\m$ backbend.  
This can be viewed as a complete straightening of the classical spiral and
happens when $\mu=\mu_{gr}\approx 2.67\times 10^3$. For $\mu<\mu_{gr}$
the system exhibits a normal first-order phase transition.  To
illustrate this, in Figs.~\ref{maxwell1}-\ref{maxwell2} we present 
entropy and inverse temperature plots for $\mu=10^3$; signs of
microcanonical first-order phase transition such as convex entropy dip
and region of negative specific heat where $d^2s/d\e^2>0$ are clearly visible.

As we decrease the degeneracy parameter $\mu$ even further, the convex
dip in the entropy vs. energy plot and correspondingly the interval
where $d^2s/d\e^2>0$ get narrower, and for $\mu=\mu_{cr} \approx 83$
they disappear (Fig.~\ref{multimu2}). At $\mu=\mu_{cr}$ the equation
$d^2s/d\e^2=0$ has only one real root, $\e=\e_{cr}\approx-0.5$, a
critical point where the two phases become indistinguishable and the heat 
capacity
diverges.  This corresponds to the line of first-order phase transitions
in $(\e,\mu)$ space, terminated by the critical point at
$(\e_{cr},\mu_{cr})$ where the phase transition is second order (see
Fig. \ref{diag}).

For $\mu<\mu_{cr}$, the inverse temperature $\b(\e)$ is a
monotonically decreasing function of energy $\e$ and the system does
not exhibit any phase transition.  Therefore, as the degeneracy
parameter decreases, the microcanonical ensemble of self-gravitating
fermions consecutively exhibits gravitational, first-order,
second-order, and no phase transition at all. Additional discussion of
phase transitions in the self-gravitating Fermi gas can be found in
Ref. \cite{ch}.

\section{self-attracting  particles with soft Coulomb potential}
\label{sec_soft}

In this section, we consider phase transitions in another
self-attracting system where the short-range cutoff is explicitly
present in the interaction potential. As in Sec.~\ref{sec_fermions},
we consider a microcanonical ensemble of $N\gg 1$ identical unit mass
particles confined to a spherical container of radius $R$, but now the
particles obey classical statistics and interact via the attracting
soft Coulomb potential $-G/(r^2+r_0^2)^{-1/2}$.  This potential has
been used in various numerical simulations of self-gravitating systems
and is also called the Plummer potential. Since such
potential does not satisfy the Poisson equation, the
equation for the mean-field density or potential cannot be reduced to a 
differential equation similar to
Eq.~(\ref{p}). In this section, we use the integral equation approach
suggested in Ref. \cite{usg}.  
The entropy $S(E)$ of the system  
is defined as the
logarithm of the density of states with the energy $E$,
\begin{eqnarray}
\label{s1}
\nonumber
S(E)  = \log\left\{ {1 \over N !}\int\ldots \int
\prod_{k=1}^M{{d^3 {\bf p}_k\, d^3 {\bf r}_k}\over{(2\pi\hbar)^3}}\right.
\\
\times \delta\!\left[E - \sum_{l=1}^M\;{{p_l^2}\over 2}+
\left.\sum_{i=1}^M \sum_{j=1+1}^M {G \over \sqrt{({\bf r}_i-{\bf
r}_j)^2+
r_0^2}} \right] \right\}.
\end{eqnarray}
Following the steps described in Refs. \cite{dv,usg}, we integrate
Eq. (\ref{s1}) on momenta, express the remaining configurational
integral through a functional integral over
possible density profiles $\r({\bf r})$, apply the saddle-point
approximation, and introduce the
dimensionless coordinate ${\bf x} \equiv {\bf r}/R$,
soft potential radius $x_0\equiv r_0/R$, and energy per particle
$\e\equiv E R/GM^2$. The saddle-point density profile
$\r_s({\bf x})$ satisfies the following integral equation
\begin{eqnarray}
\label{extr}
\r_s({\bf x})=\r_0 F[\r_s(.), {\bf x}],\qquad\qquad\qquad\nonumber\\
F[\r_s(.),{\bf x}]=\exp\left [\b \int {\r_s({\bf x}')\over
\sqrt{({\bf x}-{\bf x}')^2+x_0^2}}d^3{\bf x}'\right ],\nonumber\\
\b={3\over 2} \left[\e + {1\over 2}\int \int{\r_s({\bf x}_1) \r_s({\bf x}_2)
\over \sqrt{({\bf x}_1-{\bf x}_2)^2+x_0^2} }d^3{\bf x}_1 d^3{\bf x}_2
\right ]^{-1},\nonumber\\
\r_0=\left[\int F[\r_s(.), {\bf x}]d^3{\bf x}\right]^{-1}.\qquad\qquad\qquad
\end{eqnarray}
Neglecting an energy-independent constant, the saddle-point entropy
per particle $s(\e)\equiv S(\e)/M$ is expressed as
\be
\label{s2}
s(\e)=-{3\over 2}\ln\b-\int\r_s({\bf x})\ln[\r({\bf x})] d^3 {\bf x},
\end{equation}
where $\b$, introduced in Eq. (\ref{extr}), is the inverse
dimensionless temperature $\b(\e)=ds/d\e$. In order to solve
Eqs.~(\ref{extr}), we assume spherical symmetry of $\r$, integrate
Eqs.~(\ref{extr}) over angular variables and introduce a map
\cite{usg}:
\be
\label{it}
\r_{i+1}(x)=\s\r_0 F[\r_i(.),x]+(1-\s)\r_i(x)
\end{equation}
which we iterate numerically.
Here $F[\r(.),x]$ and $\r_0$ are defined in Eqs. (\ref{extr});
$0<\s\leq 1$ is a positive step parameter. For convergence,
the step parameter is usually set to $\s \sim 10^{-1}$.
As pointed out in Ref. \cite{usg}, the integral equation method outlined above
allows to obtain density profiles corresponding only to
thermodynamically stable or metastable states; for unstable states the
iterations defined by Eq. (\ref{it}) diverge for any $\s$.

\begin{figure}
\includegraphics[width=.45\textwidth]{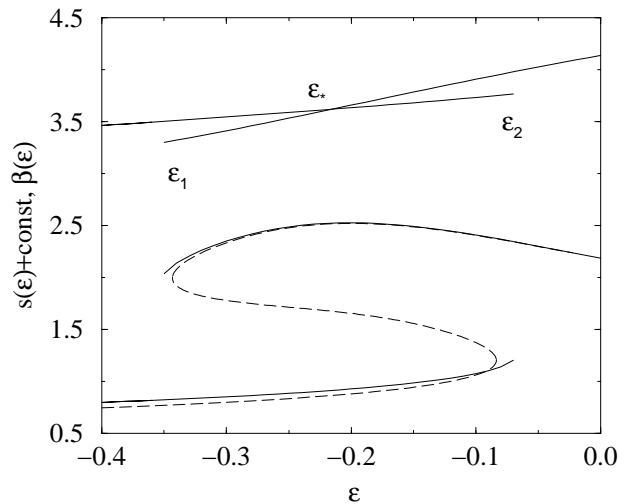}
\caption{\label{core1}
Plots of entropy $s(\e)$ (top) and inverse
temperature $\b(\e)=ds/d\e$ (bottom) vs. energy $\e$
for the soft potential radius $x_0=10^{-2}$. Plot of
 $\b(\e)$ for the fermionic system with $\mu=10^4$ is shown
in dashed line.
}
\end{figure}

\begin{figure}
\includegraphics[width=.45\textwidth]{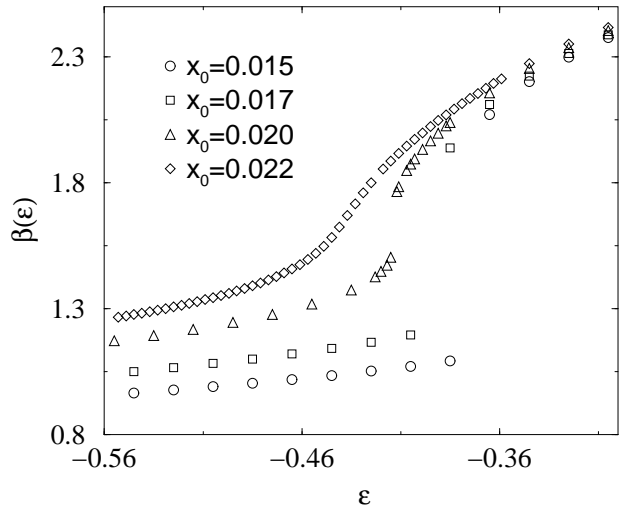}
\caption{\label{core2}
Plots of inverse temperature $\b(\e)=ds/d\e$ vs. energy $\e$ for 
different soft potential radii.}
\end{figure}

\begin{figure}
\includegraphics[width=.45\textwidth]{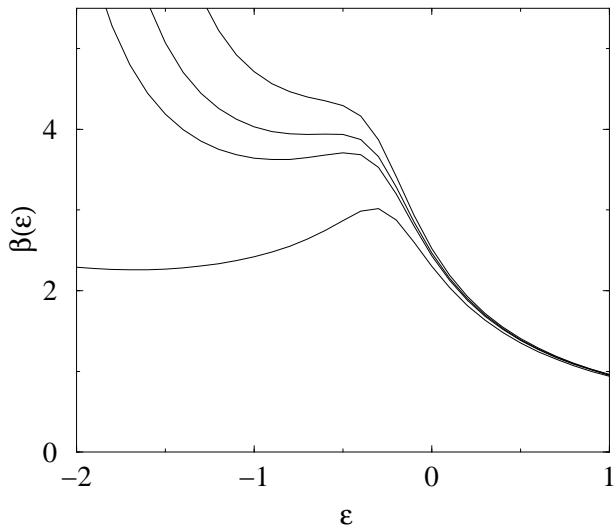}
\caption{\label{core3} Plots of inverse
temperature $\b(\e)=ds/d\e$ vs. energy $\e$
for the soft potential radii
(bottom to top) $x_0=0.12, \;
0.2, \; 0.22$, and 0.25. 
}
\end{figure}

The results in the form of $s(\e)$ vs. $\e$ and $ds/d\e$ vs. $\e$
plots for various soft potential radii $x_0$ are presented in 
Figs~\ref{core1}-\ref{core3}.
In Fig.~\ref{core1} we show the entropy and inverse temperature plots for a
relatively small soft potential radius, $x_0=10^{-2}$. For this value
of $x_0$ the system clearly exhibits all signs of gravitational phase
transition.  For comparison, in the same figure we present the
$\b(\e)$ plot for a low-degeneracy fermionic system with $\mu=10^4$
from Fig.~3.  Despite the completely different nature of the short-range
cutoffs for these systems, their uniform state entropies exhibit a
strikingly similar behavior.  This once again illustrates that
properties of a uniform state are determined mostly by the long-range
properties of the interaction.  Naturally, the core-halo state structure
and its properties depend on the nature of the cutoff, so the
corresponding branches in the entropy vs. energy plot are visually
different. Nevertheless, the difference is weak so the physical properties
of phase transitions in long-range self-attracting systems are relatively
insensitive to the precise form of the small-scale regularization.

Similarly to the fermionic system, as the range of the cutoff
is increased, the range of existence of metastable states
shrinks and finally disappears. At this point the gravitational phase
transition crosses over to the first-order one. From the data presented in
Fig.~\ref{core2} we estimate that this crossover happens at $x_0=x_{gr}\approx0.021$.
For $x_{gr}<x_0<x_{cr}$ the system exhibits a normal first-order phase
transition until the critical point $(\e_{cr},x_{cr})$ is reached.
The plots presented in Fig.~\ref{core3} indicate that $\e_{cr}\approx-0.7$ and
$x_{cr}\approx 0.22$. When $x_0>x_{cr}$, no phase transitions are
present in the system.
This allows us to conclude that similarly to the fermionic system, as the 
soft potential
radius is increased, the self-attracting system with soft Coulomb
interaction exhibits consecutively gravitational, first-order,
second-order, and no phase transition at all.

\section{validity of mean field approximation}
\label{sec_mf}

To obtain the plots shown in Figs.~\ref{fignew}-\ref{core3} we used the
saddle point or the mean-field approximation. It raises an important
question of whether the distinct features of the gravitational phase
transition described above are intrinsic or appear as artifacts of
this approximation.  Before the approximation is applied, the
microcanonical entropy $S(E,N)$ is expressed through the logarithm of
a sum of microscopic densities of states $W_i$ of all macroscopic
states with the energy $E$ and number of particles $N$ ($k_B=1$).  The
mean-field approximation is equivalent to replacing a sum of
contributions from these macroscopic states, usually represented by a
functional integral over corresponding density (or phase space density)
profiles $\r$, by a
contribution from the single state or density profile $\r_0$  \cite{hk,dv,pr}:
\begin{equation}
\label{h}
S=\ln \sum_i W_i \sim \ln \int {\cal D} \r \ {W}[\r]\approx \ln {W}[\r_0].
\end{equation}
This state $\r_0$ extremizes the density of states $W$ and, consequently, the 
entropy $S$. The condition
\begin{equation}
\left.{\d W \over \d \r}\right|_{\r=\r_0}=0,
\end{equation}
defines $\r_0$ that gives a global maximum, a local maximum, or a
minimum (or saddle point) to the entropy which correspond to stable,
metastable, or unstable states, respectively. Let us first consider
the range of energies, ($E<E_1$ and $E>E_2$ in
Figs.~\ref{EbetaSMU10e4} and \ref{core1}, or all energies in
Figs.~\ref{multimu2} and \ref{core3}, for which there exists only one
global entropy maximum. This case is described in traditional
textbooks of thermodynamics: the single equilibrium state corresponds
to a very sharp maximum in the density of states and the first-order
corrections to the entropy per particle $s$ scales as $1/N$,
i.e. $s=s[\r_0]+{\cal O}(1/N)$.  When stable and metastable states
coexist ($E_1<E<E_2$), the mean-field approximation is also an
asymptotically exact approximation for the stable states, since the
relative contributions from the metastable states $\r_m$ to ${W}[\r]$
scales as $\exp\{N(s[\r_m]-s[\r_0])\}$ However, the sharp kink in the
mean-field entropy plot at $E^*$ (as in Fig.~\ref{ESintro}) appears in
the true, non-mean-field entropy plot only in the $N\rightarrow\infty$
limit; as for any finite number of particles the metastable states
contribute significantly to ${W}[\r]$ in the vicinity of $E^*$, where
$s[\r_m]-s[\r_0]\rightarrow 0$.  Similarly, the mean-field
approximation works well for the metastable states when they are sharp
local maxima of ${W}[\r]$.  But this approximation breaks down when
the contributions to the entropy from the metastable state $\r_m$
becomes comparable to or less than the contributions from other states
$\r'$ in a vicinity of $\r_m$, $\|\r_m-\r'\| \ll \|\r_m\|$.  This
happens when $\r_m$ ceases to be at least a local maximum of entropy,
which is exactly what takes place at the metastability-instability
transition points $E_1$ and $E_2$.  This breakdown of the mean field
approximation near the $E_1$ and $E_2$ energies can also be viewed as
fluctuation-induced uncertainty in the exact location of the
metastability-instability transition. It is shown in Ref. \cite{ka2}
that the relative uncertainty $\Delta E/E_1$ in the position of the
collapse point $E_1$ scales with the number of particles as
$N^{-2/3}$.  Hence, given that $N$ is large, the mean-field results
are asymptotically exact for all energies except for those near the
ends of metastable branches $E_1$ and $E_2$.

Another distinct feature of gravitational phase transitions is the anomalous
stability of the metastable branches $[E_1,E^*]$ and $[E^*,E_2]$
(Fig.~\ref{ESintro}). Consider, for example, a metastable uniform
state and a stable core-halo state both having the same energy
somewhere in the middle of the interval $[E_1,E^*]$. The entropy
minimum, that separates the entropy maxima corresponding to the stable
and metastable states, has the depth $\D S$ which is proportional to
the number of particles, i.e. $\D S=N\D s$ (for example, in
Fig.~\ref{EbetaSMU10e4} where $\D s$ is equal to the difference in
coordinates between the metastable (solid line) and unstable (dashed
line) states, $\D s\approx 0.1$). Physically, this is so because the
transition from a metastable uniform state to a stable core-halo state
requires a macroscopic fluctuation equivalent to the rearrangement of
the density distribution in the {\it whole} system.  Hence, the
probability of the metastable-stable transition is proportional to
$\exp(-N\D s)$ and becomes prohibitively small even for a moderate
number of particles $N$.  Only near the ends of metastability branches
$E_1$ and $E_2$ is the probability of metastability-stability
transition significant; it is of order ${\cal O}(N^0)$ in the interval
$[E_1, E_1+\Delta E]$, where $\Delta E\sim E_1 N^{-2/3}$ \cite{ka2}.

Given the arguments presented above we conclude that
the mean-field approximation adequately represents the
phenomenology of self-attracting systems and correctly describes
the gravitational, first-order, and second-order
phase transitions.

\section{conclusion}
\label{sec_conclusion}

\begin{figure}
\includegraphics[width=.45\textwidth]{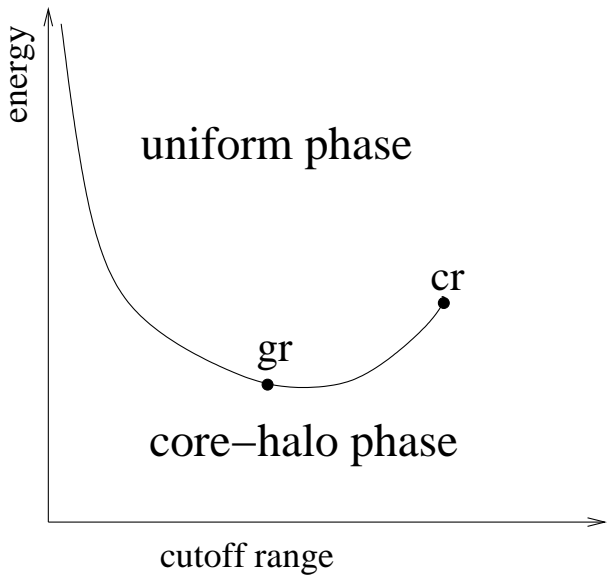}
\caption{\label{diag} Sketch of the phase diagram
of a self-attracting system. The crossover point between
the gravitational and first-order phase transition and the critical point
are marked ``gr'' and ``cr'', respectively.
}
\end{figure}

In Sect. \ref{sec_fermions} and \ref{sec_soft} we considered two
examples of self-attracting systems, the ensemble of self-gravitating
fermions and the ensemble of classical particles interacting via
attractive soft Coulomb potential. These systems have a similar $\sim
1/r$ interaction at large distances but very different forms of
short-range cutoffs.  While in the second example the short-range
cutoff is evidently $r_0$, in the first example the role of the
short-range cutoff is indirectly played by the Pauli exclusion
principle, which depends on the particle density.  Despite this
different small-$r$ behavior, both of the considered ensembles exhibit
the same sequence of phase transitions: gravitational, first-order,
second-order, and none, as the range of their respective cutoffs is
increased.  The sketch of their phase diagram in cutoff-energy
coordinates is represented in Fig.~\ref{diag} and completes the one given in
Ref. \cite{ch} in cutoff-temperature coordinates.
The examples considered in this paper were chosen mainly because of
the physical importance of the $1/r$ potential but similar phase diagrams
exist for all nonintegrable $1/r^{\a}$, $0<\a<3$ attractive
potentials.  The main physical reason behind this phase diagram is
that the short-range cutoff controls the maximum density a
self-attracting system can achieve.  As the range of the cutoff is
increased, the collapsed central core becomes less dense and occupies
more volume, and at some point the system has simply nowhere to
collapse. It happens when the central density of the non-collapsed uniform 
state
becomes similar to the core density of the collapsed core-halo
state. Likewise, the critical point is reached when the maximum
allowed density becomes so small that the system remains virtually
uniform for any energy.  The purely geometrical nature of these
arguments indicates that the phase diagram, obtained in the previous
two sections, should be robust and insensitive to such simplifying
assumptions as spherical symmetry.  The validity of the main
approximation used in this paper, the mean-field approach, was
discussed in Sec. \ref{sec_mf}.  It is revealed that the mean-field
approximation correctly describes the behavior of the self-attracting
systems for all accessible energies  and cutoff radii, excluding the
immediate vicinities of the collapse points.  We leave the study of
the collapse points in the finite-particle systems 
as well as the dynamics of the collapse for a
future paper.

\section{acknowledgment}

The authors are grateful to the organizers of the Les Houches Ecole de
Physique meeting for providing an opportunity to discuss
the results of this work. I.~I. gratefully acknowledges the support
of  FONDECYT under grant 1020052.

\end{document}